\documentclass{ws-ijmpc}
\usepackage{latexsym}
\usepackage[lined,algonl,boxed]{algorithm2e}
\usepackage{tipa}

\begin{document}

\markboth{Mukherjee \bgroup et al. \egroup}
{Modeling the Co-occurrence Principles of Consonant Inventories}

\catchline{}{}{}{}{}

\title{Modeling the Co-occurrence Principles of the Consonant Inventories:\\ A Complex Network Approach}

\author{Animesh Mukherjee, Monojit Choudhury, Anupam Basu \and Niloy Ganguly}

\address{Department of Computer Science and Engineering\\ Indian Institute of Technology, Kharagpur--721302\\ 
\{animeshm, monojit, anupam, niloy\}@cse.iitkgp.ernet.in}

\maketitle

\begin{history}
\received{Day Month Year}
\revised{Day Month Year}
\end{history}

\begin{abstract}
Speech sounds of the languages all over the world show remarkable patterns of co-occurrence. In this work, we attempt 
to automatically capture the patterns of co-occurrence of the consonants across languages and at the same time figure 
out the nature of the force leading to the emergence of such patterns. For this purpose we define a weighted network 
where the consonants are the nodes and an edge between two nodes (read consonants) signify their co-occurrence 
likelihood over the consonant inventories. Through this network we identify communities of consonants that essentially 
reflect their patterns of co-occurrence across languages. We test the goodness of the communities and observe that the 
constituent consonants frequently occur in such groups in real languages also. Interestingly, the consonants forming 
these communities reflect strong correlations in terms of their features, which indicate that the principle of feature 
economy acts as a driving force towards community formation. In order to measure the strength of this force we propose 
an information theoretic definition of feature economy and show that indeed the feature economy exhibited by the 
consonant communities are substantially better than those if the consonant inventories had evolved just by chance.

\end{abstract}

\ccode{PACS Nos.: 89.75.-k, 89.75.Fb}

\section{Introduction}\label{intro}

Sound inventories of the world's languages show remarkable regularities. Any randomly chosen set of consonants and 
vowels does not make up the sound inventory of a particular language. In fact one of the earliest observations about 
the consonant inventories has been that consonants tend to occur in pairs that exhibit strong correlation in terms of 
their {\em features}\footnote{In linguistics, features are the elements, which distinguish one phoneme from another. 
The features that distinguish the phonemes can be broadly categorized into three different classes namely the {\em 
manner of articulation}, the {\em place of articulation} and {\em phonation}. Manner of articulation specifies how the 
flow of air takes place in the vocal tract during articulation of a phoneme whereas place of articulation specifies 
the active speech organ and also the place where it acts. Phonation describes the activity regarding the vibration of 
the vocal cords during the articulation of a phoneme.}~\cite{Trub:30}. In other words, consonants have a tendency to 
form groups or communities that effectively reflect their patterns of co-occurrence across the languages of the world. 
In order to explain these trends, {\em feature economy} was proposed as the basic organizing principle of the 
consonant inventories~\cite{Groot:31,Martinet:55}. According to this principle, languages tend to maximize the 
combinatorial possibilities of a few distinctive features to generate a large number of consonants. Stated 
differently, a given consonant will have a higher than expected frequency in inventories in which all of its features 
have distinctively occurred in other sounds. The idea is illustrated, with an example, through Table~\ref{tab1}. 
Although there have been several attempts to explain the observed co-occurrence patterns~\cite{Clements:04} through 
linguistic insights~\cite{Boersma:98}, as far as our knowledge goes there has been no work to identify the communities 
of consonants algorithmically.    

\begin{table}\centering
\tbl{The table shows four plosives two of which are voiced, and, the other two are voiceless. It also indicates the 
two different places of articulation (dental and bilabial) for these plosives. If a language has in its consonant 
inventory any three of the four entries of this table, then there is a higher than average chance that it will also 
have the fourth entry of the table in its inventory.}
{\begin{tabular}{|l|l|l|}
\cline{1-3}
\vbox to2.04ex{\vspace{1pt}\vfil\hbox to12.80ex{\hfil plosive \hfil}} & 
\vbox to2.04ex{\vspace{1pt}\vfil\hbox to12.80ex{\hfil voiced\hfil}} & 
\vbox to2.04ex{\vspace{1pt}\vfil\hbox to12.80ex{\hfil voiceless\hfil}} \\

\cline{1-3}
\vbox to1.88ex{\vspace{1pt}\vfil\hbox to12.80ex{\hfil dental\hfil}} & 
\vbox to1.88ex{\vspace{1pt}\vfil\hbox to12.80ex{\hfil /$d$/\hfil}} & 
\vbox to1.88ex{\vspace{1pt}\vfil\hbox to12.80ex{\hfil /$t$/\hfil}} \\

\cline{1-3}
\vbox to1.88ex{\vspace{1pt}\vfil\hbox to12.80ex{\hfil bilabial\hfil}} & 
\vbox to1.88ex{\vspace{1pt}\vfil\hbox to12.80ex{\hfil /$b$/\hfil}} & 
\vbox to1.88ex{\vspace{1pt}\vfil\hbox to12.80ex{\hfil /$p$/\hfil}} \\

\cline{1-3}
\end{tabular}
\label{tab1}}
\end{table}

In this work, we propose {\em a  method to automatically capture the patterns of co-occurrence of the consonants 
across languages} and at the same time {\em quantify the driving force leading to the emergence of such patterns.} For 
this purpose, we define the ``Phoneme-Phoneme Network" or {\bf PhoNet}, which is a weighted network where the 
consonants are the nodes and an edge between two nodes (read consonants) signify their co-occurrence likelihood over 
the consonant inventories. We conduct empirical studies of PhoNet and analyze it from the perspective of a social 
network where consonants exhibit community structures. Recently, several complex phenomena observed in the social, 
biological and physical worlds have been modeled as networks, which provides a comprehensive view of their underlying 
organizational principles. See~\cite{Albert:02,Newman:03} for a review on modeling and analysis of such networked 
systems. There have been some attempts as well to model the intricacies of human languages through complex networks. 
Word networks based on synonymy~\cite{Yook:01}, co-occurrence~\cite{Cancho:01}, and phonemic 
edit-distance~\cite{Vitevitch:05} are examples of such attempts. As a matter of fact, the distribution of the 
consonants across languages have also been modeled as a complex bipartite network in~\cite{Choudhury:06}, but the 
study is limited to the occurrence of the consonants and not their co-occurrence.  

This article is organized as follows: Section~\ref{ph_def} formally defines PhoNet and outlines its construction 
procedure. In section~\ref{comm} we employ the extended Radicchi \bgroup et al. \egroup~\cite{Rad:03} algorithm, to 
find the communities in PhoNet. In section~\ref{stat} we test the goodness of the communities and observe that the 
constituent consonants of these communities frequently occur in such groups in real languages also. Interestingly, the 
consonants forming these communities reflect strong correlations in terms of their features, which points to the fact 
that feature economy binds these communities. In order to quantify feature economy we propose an information theoretic 
approach in section~\ref{fe}. In the same section we show that the feature economy exhibited by the consonant 
communities obtained from PhoNet are indeed substantially better than those, if the consonant inventories had evolved 
just by chance. We also show that the number of languages in which the consonants of a community occur together 
increases with increasing feature economy. Finally we conclude in section~\ref{conc} by summarizing our contributions, 
pointing out some of the implications of the current work and indicating the possible future directions.

\section{PhoNet: The Phoneme-Phoneme Network}\label{ph_def}

\begin{figure}
\framebox{
\centerline{\psfig{file=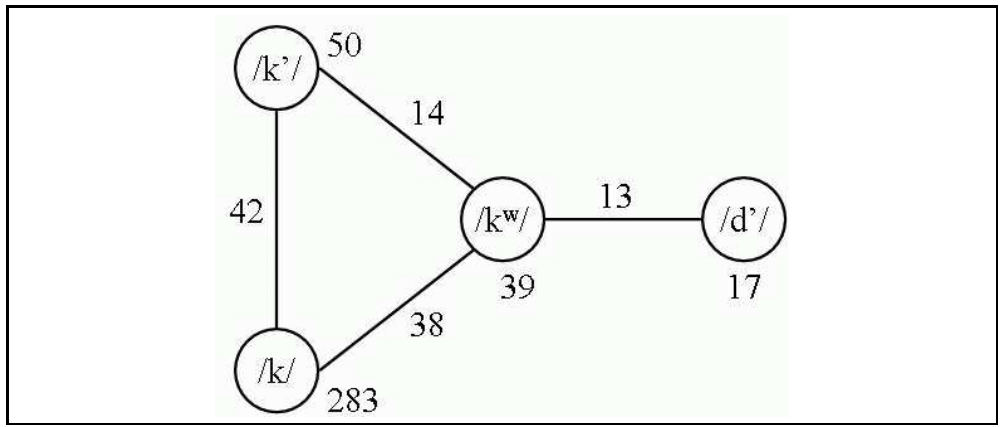,width=3in}}
}
\caption{A partial illustration of the nodes and edges in PhoNet. The labels of the nodes denote the consonants 
represented in IPA (International Phonetic Alphabet). The numerical values against the edges and nodes represent their 
corresponding weights. For example /$k$/ occurs in 283 languages; /$k^w$/ occurs in 39 languages while they co-occur 
in 38 languages.}
\label{graph}
\end{figure}

We define PhoNet as a network of consonants, represented as G = $\langle$ V$_C$, E $\rangle$ where V$_C$ is the set of 
nodes labeled by the consonants and E is the set of edges occurring in PhoNet. There is an edge $e$ $\in$ E between 
two nodes, if and only if there exists one or more language(s) where the nodes (read consonants) co-occur. The weight 
of the edge $e$ (also {\em edge-weight}) is the number of languages in which the consonants connected by $e$ co-occur. 
The weight of a node $u$ (also {\em node-weight}) is the number of languages in which the consonant represented by $u$ 
occurs. In other words, if a  consonant $c_i$ represented by the node $u$ occurs in the inventory of $n$ languages 
then the node-weight of $u$ is assigned the value $n$. Also if the consonant $c_j$ is represented by the node $v$ and 
there are $w$ languages in which consonants $c_i$ and $c_j$ occur together then the weight of the edge connecting $u$ 
and $v$ is assigned the value $w$. Figure~\ref{graph} illustrates this structure by reproducing some of the nodes and 
edges of PhoNet.  

\subsection{Construction of PhoNet}

\begin{figure*}
\centerline{\psfig{file=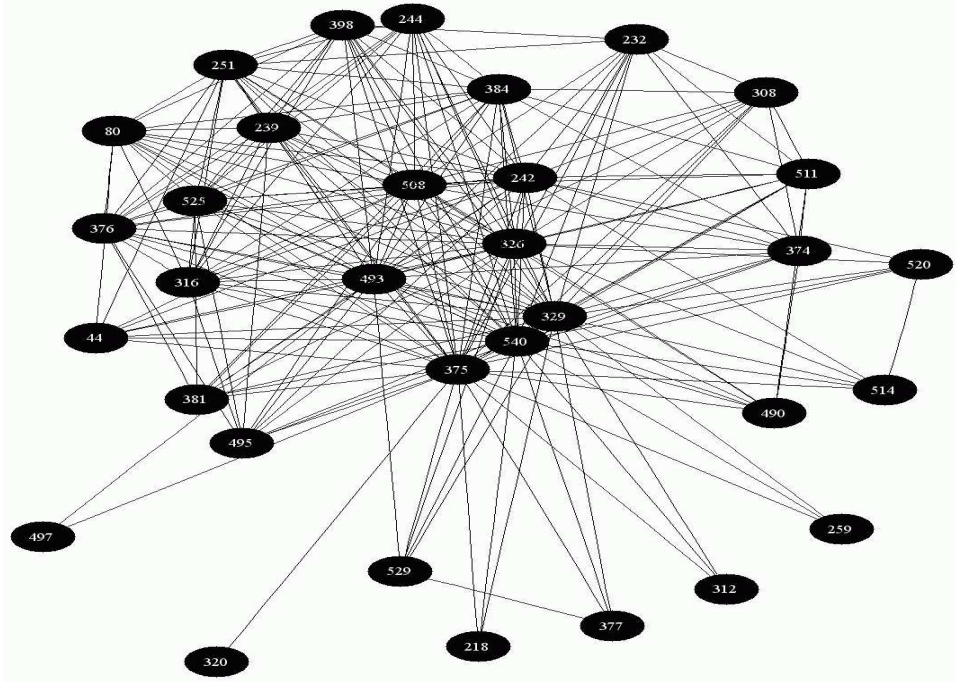,width=5in}}
\caption{A partial illustration of PhoNet. All edges in this figure have an edge-weight greater than or equal to 50. 
The number on each node corresponds to a particular consonant. For instance, node number 508 corresponds to /$g$/ 
whereas node number 540 represents /$k$/.}
\label{ph10}
\end{figure*}

Many typological studies~\cite{Hinskens:03,Ladefoged:96,Lindblom:88} of segmental inventories have been carried out in 
past on the UCLA Phonological Segment Inventory Database (UPSID)~\cite{Maddieson:84}. UPSID initially had 317 
languages and was later extended to include 451 languages covering all the major language families of the world. In 
this work we have used the older version of UPSID comprising of 317 languages and 541 consonants (henceforth 
UPSID$_{317}$), for constructing PhoNet. Consequently, the set V$_C$ comprises of 541 elements (nodes) and the set E 
comprises of 34012 elements (edges). At this point it is important to mention that in order to avoid any confusion in 
the construction of PhoNet, we have appropriately filtered out the {\em anomalous} and the \emph{ambiguous} 
segments~\cite{Maddieson:84} from it. In UPSID, a segment has been classified as anomalous if any of the following 
conditions holds: the segment is (1) rare (very low frequency), (2) occurs only in loans, (3) is existent only in 
underlying forms, (4) is derivable from other segments, or (5) obscure in description. We have completely ignored the 
anomalous segments from the data set. Ambiguous segments are those for which UPSID provides insufficient information. 
For example, the presence of both the palatalized dental plosive and the palatalized alveolar plosive are represented 
in UPSID as palatalized dental-alveolar plosive. In absence of any descriptive sources explaining how such ambiguities 
might be resolved, we have decided to include them as distinct segments. A similar treatment of anomalous and 
ambiguous segments has also been described in Pericliev and Vald{\'e}s-P{\'e}rez~\cite{Peri:02}. Figure~\ref{ph10} 
presents a partial illustration of PhoNet as constructed from UPSID$_{317}$.

\section{Identification of Community Structures}\label{comm}

There is a large volume of literature suggested by computer scientists, physicists as well as sociologists that speaks 
about identifying communities in a network~\cite{Flake:02,Holme:03,Ker:70,Newman:04,Pothen:90,Rad:03}. This is mainly 
because, the ability to find communities within large networks in some automated fashion could be of considerable use.
Communities in a web graph for instance might correspond to sets of web sites dealing with related 
topics~\cite{Flake:02}, while communities in a biochemical network might correspond to functional units of some kind 
~\cite{Holme:03}. 

In this work we attempt to identify the communities appearing in PhoNet by extending the Radicchi \bgroup et al. 
\egroup~\cite{Rad:03} algorithm for weighted networks\footnote{We have tried a few other community finding algorithms 
but this algorithm performs slightly better. Moreover, we found it easier to extend this algorithm to weighted 
networks.}. The algorithm of Radicchi \bgroup et al. \egroup (applied on unweighted networks) counts, for each edge, 
the number of loops of length three it is a part of and declares that edges with very low counts as inter-community 
edges. 
\paragraph{{\bf Basis:}} Edges that run between communities are unlikely to belong to many short loops, because to 
complete a loop containing such an edge there needs to be another edge that runs between the same two communities, and 
such other edges are rare. 
\paragraph{{\bf Modification for Weighted Network:}} Nevertheless, for weighted networks, rather than considering 
simply the triangles (loops of length three) we need to consider the weights on the edges forming these triangles. The 
basic idea is that if the weights on the edges forming a triangle are comparable then the group of consonants 
represented by this triangle highly occur together rendering a pattern of co-occurrence while if these weights are not 
comparable then there is no such pattern. In order to capture this property we define a strength metric $S$ for each 
of the edges of PhoNet as follows. Let the weight of the edge ($u$,$v$), where $u$, $v$ $\in$ V$_C$, be denoted by 
$w_{uv}$. We define $S$ as,         
\begin{equation}
S =  \frac{w_{uv}}{\sqrt{\sum_{i \in V_C - \{u,v\}}{(w_{ui} - w_{vi})}^2}}    
\end{equation} 
if $\sqrt{\sum_{i \in V_C - \{u,v\}}{(w_{ui} - w_{vi})}^2} > 0$ else $S = \infty$. The denominator  in this expression 
essentially tries to capture whether or not the weights on the edges forming triangles are comparable. If the weights 
are not comparable then this denominator will be high, thus reducing the overall value of $S$ . PhoNet may be then 
partitioned into clusters or communities by removing edges that have $S$ close to zero. 

\begin{figure}[h]
\framebox{
\centerline{\psfig{file=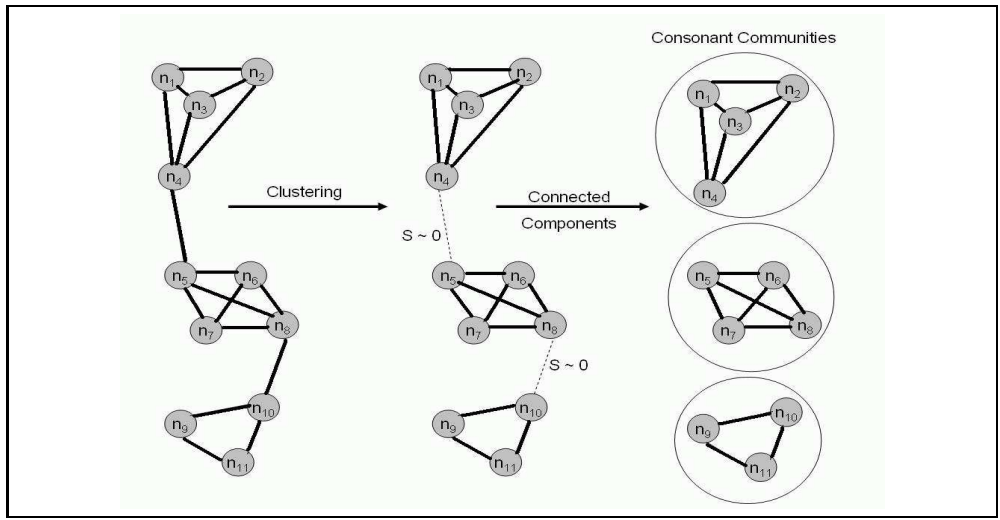,width=4in}}
}
\caption{The process of community formation}
\label{cluster_process}
\end{figure}

\begin{figure}[h]
\framebox{
\centerline{\psfig{file=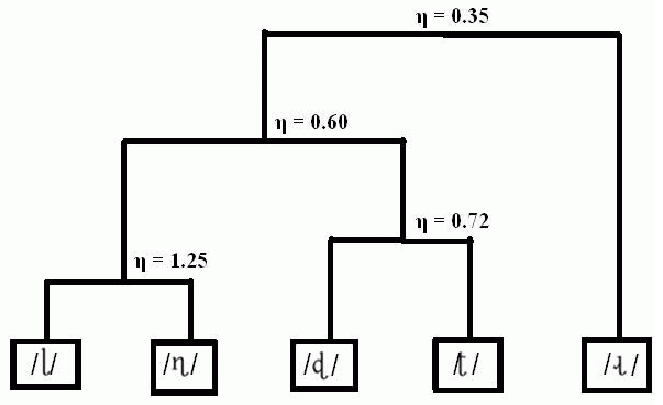,width=3in}}
}
\caption{The dendrogram illustrates how the retroflex community of \textipa{/\:d/, /\:t/, /\:n/, /\:l/} and 
\textipa{/\:R/} is formed with the change in the value of $\eta$}
\label{tree}
\end{figure}

In this algorithm we have neglected the edges in PhoNet that are connected to nodes having very low or very high 
node-weights since they are either insignificant\footnote{We have neglected nodes with node-weight less than 5 since 
these nodes correspond to consonants that occur in less than 5 languages in UPSID$_{317}$ and the communities they 
form are therefore statistically insignificant.} or assortative\footnote{We have neglected nodes with node-weight 
greater than 130. These nodes correspond to consonants that occur in more than 130 languages in UPSID$_{317}$ and 
therefore they co-occur with almost every other consonant. Hence the strength metric $S$ is likely to be high for an 
edge connecting nodes (read consonants) with high node-weights. This edge (owing to its high strength) might then 
force two otherwise disjoint communities to form a single community. For instance, we have observed that since the 
consonants /$m$/ and /$k$/ are very frequent, the nodes corresponding to both of them have a high node-weight and 
consequently the edge between them also has a high edge-weight. The strong link between /$m$/ and /$k$/ then forces 
the respective bilabial and velar communities to merge into a single community.} (see ~\cite{Newman:03} for a 
reference) respectively. Henceforth we will refer to this version of PhoNet as PhoNet$_{red}$. The entire idea is 
summarized in Algorithm~\ref{wght}. Figure~\ref{cluster_process} illustrates the clustering process. We can obtain 
different sets of communities by varying the threshold $\eta$. As the value of $\eta$ decreases, new nodes keep 
joining the communities and the process is similar to hierarchical clustering~\cite{Scott:00}. Figure~\ref{tree} shows 
a dendrogram, which illustrates the formation of the community of the consonants \textipa{/\:d/, /\:t/, /\:n/, /\:l/} 
and \textipa{/\:R/} with the change in the value of $\eta$.

\protect\begin{algorithm}
\BlankLine
\KwIn{PhoNet$_{red}$}
 \Repeat{Phonet$_\eta$ gets fully connected}
 {
  \For{each edge (u,v)}
  {
   Compute \\
	$S = \frac{w_{uv}}{\sqrt{\sum_{i \in V_C - \{u,v\}}{(w_{ui} - w_{vi})}^2}}$  \\
	if $\sqrt{\sum_{i \in V_C - \{u,v\}}{(w_{ui} - w_{vi})}^2} > 0$ else $S = \infty$;
  }
  \BlankLine
  Redefine the edge-weight for each edge ($u$,$v$) by $S$;
  \BlankLine
  Remove edges with edge-weights less than or equal to a threshold $\eta$;\\ 
  Call this new version of PhoNet, PhoNet$_\eta$;
  \BlankLine
  Find the connected components in PhoNet$_\eta$;
  \BlankLine
  $\eta = \frac{\eta}{\delta}$ where $\delta$ is the diminishing factor;
 }  
\label{wght}
\caption{Algorithm for finding communities based on edge strength}    
\end{algorithm}

Some of the example communities obtained from our algorithm are noted in Table~\ref{communities}. In this table, the 
consonants in the first community are dentals, those in the second community are retroflexes, while the ones in the 
third are all laryngealized.    

\begin{table}\centering
\tbl{Consonant communities}
{\begin{tabular}{|l|l|}
\cline{1-2}
\vbox to1.97ex{\vspace{1pt}\vfil\hbox to33.00ex{\hfil Community\hfil}\vfil} & 
\vbox to1.97ex{\vspace{1pt}\vfil\hbox to21.20ex{\hfil Features in Common\hfil}} \\

\cline{1-2}
\vbox to1.70ex{\vspace{1pt}\vfil\hbox to33.00ex{\hfil /t/, /d/, /n/\hfil}\vfil} & 
\vbox to1.70ex{\vspace{1pt}\vfil\hbox to21.20ex{\hfil dental\hfil}\vfil} \\

\cline{1-2}
\vbox to1.70ex{\vspace{1pt}\vfil\hbox to33.00ex{\hfil \textipa{/\:d/, /\:t/, /\:n/, /\:l/, /\:R/}\hfil}\vfil} & 
\vbox to1.70ex{\vspace{1pt}\vfil\hbox to21.20ex{\hfil retroflex \hfil}\vfil} \\

\cline{1-2}
\vbox to1.70ex{\vspace{1pt}\vfil\hbox to33.00ex{\hfil \textipa{/\~*w/, /\~*j/, /\~*m/}\hfil}\vfil} & 
\vbox to1.70ex{\vspace{1pt}\vfil\hbox to21.20ex{\hfil laryngealized \hfil}\vfil} \\

\cline{1-2}
\end{tabular}
\label{communities}}
\end{table}

\section{Evaluation of the Communities based on their Occurrence in Languages}\label{stat}

In the earlier section we have mainly described the methods of extracting the consonant communities from PhoNet. In 
this section we look into the languages included in UPSID$_{317}$ and inspect whether or not the consonants forming 
the communities in PhoNet actually occur in such groups. 

\begin{figure*}
\centerline{\psfig{file=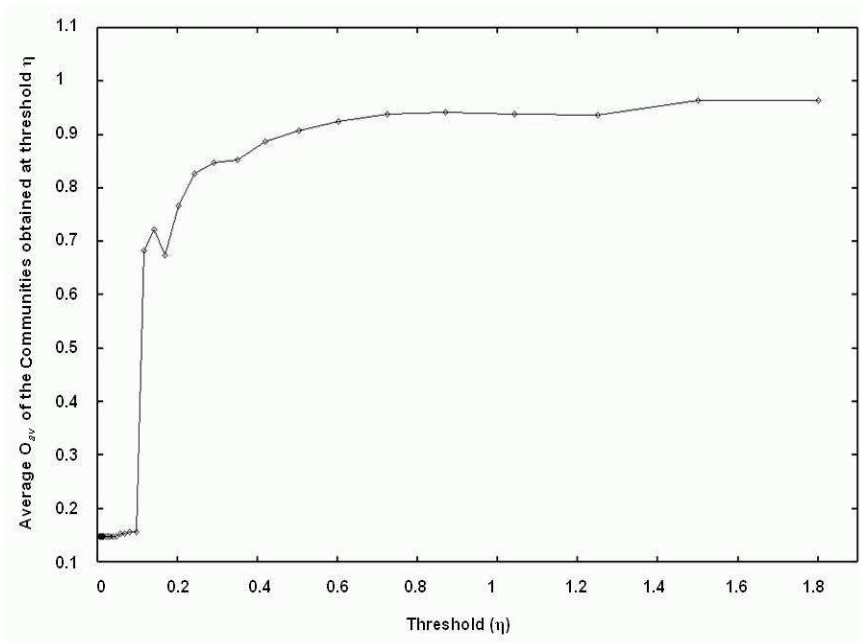,width=4.5in}}
\caption{Average $O_{av}$ of the communities obtained at a particular threshold $\eta$ versus the threshold $\eta$} 
\label{comm_eval}
\end{figure*}

For this purpose we first arrange the consonants forming a community $C$, of size $N$, in an ascending order of their 
frequency of occurrence in UPSID$_{317}$. We associate a rank $R$ with each of the consonants in $C$ where the least 
frequency consonant gets a rank $R$ = 1, the second least gets a rank $R$ = 2 and so on. Starting from rank $R$ = 1 we 
count how many of the consonants in $C$, occur in a language $L$ $\in$ UPSID$_{317}$. Let the number of such 
consonants be $M$. We define the {\em occurrence ratio} $O_L$ of the community $C$ for the language $L$ to be 
\begin{equation}
O_L = \frac{M}{N-(R_{top}-1)}
\end{equation} 
where $R_{top}$ is the rank of the highest ranking consonant that is found in $L$. The denominator of this ratio is 
$N-(R_{top}-1)$ instead of $N$ since it is not mandatory for a language to have a low frequency member of a community 
if it has the high frequency member; nevertheless if the language already has the low frequency member of the 
community then it is highly expected to also have the high frequency member\footnote{For instance let the community 
$C$ be formed of the consonants /$k^w$/, /$k^h$/ and /$k$/ as shown in Figure~\ref{graph}. When we inspect the 
language $L$ it is not necessary for it to have /$k^w$/ or /$k^h$/ if it has /$k$/ in its inventory; nevertheless it 
is highly expected that if it already has /$k^w$/, it should also have /$k$/ and /$k^h$/ in its 
inventory.}~\cite{Clements:04}.   
 The average occurrence ratio $O_{av}$ for the community $C$ can be obtained as follows,
\begin{equation}
O_{av} = \frac {\sum_{L \in UPSID_{317}}{O_L}} {L_{occur}}
\end{equation} 
where $L_{occur}$ is the number of languages in UPSID$_{317}$ that have at least one or more consonants occurring in 
$C$. Figure~\ref{comm_eval} shows the average $O_{av}$ of the communities obtained at a particular threshold $\eta$ 
versus the threshold $\eta$. The curve clearly shows that the average $O_{av}$ of the communities obtained from our 
algorithm for $\eta >$ 0.3 is always more than 0.8. This in turn implies that on an average the communities, obtained 
at thresholds above 0.3, occur in more than 80\%\footnote{The expectation that a randomly chosen set of consonants 
representing a community of size between 2 to 5, occurs in a language, is 70\%  whereas the same is 89\% for the 
communities observed in PhoNet.} of the languages in UPSID$_{317}$. At thresholds below 0.3 the average $O_{av}$ falls 
gradually since giant components start forming and the probability of all the consonants in the giant component 
occurring together in languages is very low. Hence the community structures obtained from our algorithm are true 
representatives of the patterns of co-occurrence of the consonants across languages.   

\section{Feature Economy: The Binding Force of the Communities}\label{fe}

In the earlier sections we have mainly focused ourselves to the detection and evaluation of the communities emerging 
from PhoNet. In this section we attempt to explore whether or not the driving force, which leads to the emergence of 
these communities, is feature economy. For this reason we introduce a quantitative measure of feature economy. The 
basic idea is borrowed from the concept of entropy in information theory~\cite{Shan:49}. 

\begin{figure*}
\framebox{
\centerline{\psfig{file=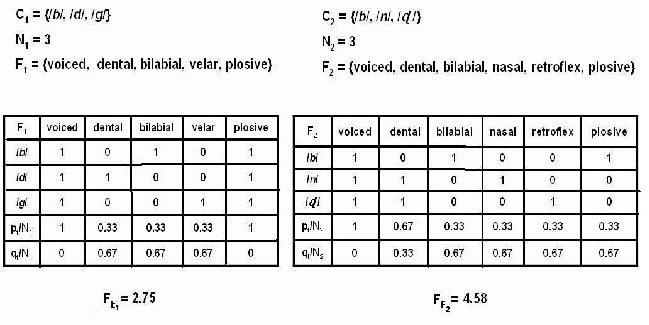,width=4.5in}}
}
\caption{The process of computing the value of $F_E$ for the two different communities $C_1$ and $C_2$}
\label{example_ent}
\end{figure*}

For a community $C$ of size $N$ let there be $p_f$ consonants, which have a particular feature $f$ (where $f$ is 
assumed to be boolean in nature) in common and $q_f$ other consonants, which lack the feature $f$. Thus the 
probability that a particular consonant chosen uniformly at random from C has the feature $f$ is $\frac {p_f}{N}$ and 
the probability that the consonant lacks the feature $f$ is $\frac {q_f}{N}$ (=1--$\frac{p_f}{N}$). If $F$ be the set 
of all features present in the consonants in C then {\em feature entropy} $F_E$ can be defined as
\begin{equation}
F_E = \sum_{f \in F}(- \frac {p_f}{N}\log{\frac{p_f}{N}} - \frac{q_f}{N}\log{\frac {q_f}{N}})        
\end{equation} 
The process of computing the values of $F_E$ for two different communities of consonants is illustrated in 
Figure~\ref{example_ent}. 

$F_E$ is essentially the measure of the minimum number of bits that are required to communicate the information about 
the entire community $C$ through a channel. Thus, the lower the value of $F_E$, the better it is in terms of 
information transmission overhead. To have more information conveyed using a fewer number of bits, maximization of the 
combinatorial possibilities of the features used by the constituent consonants in the community $C$ is needed. This is 
precisely the prediction made by the principle of feature economy\footnote{The lower the feature entropy the higher is 
the feature economy.}. In fact, it is due to this reason that in Figure~\ref{example_ent}, $F_E$ exhibited by the 
community $C_1$ is better than that of the community $C_2$, since in $C_1$ the combinatorial possibilities of the 
features is better utilized by the consonants than in $C_2$.    

\begin{figure*}
\centerline{\psfig{file=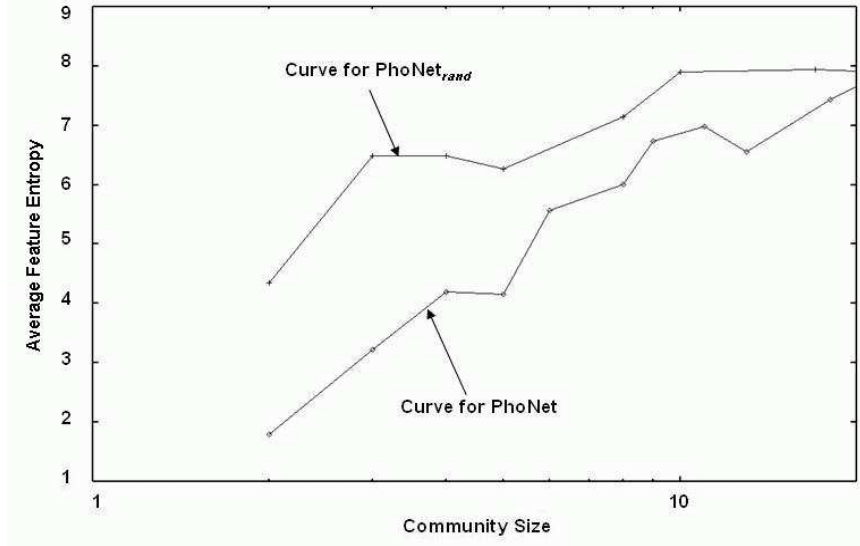,width=4.5in}}
\caption{Curves showing average feature entropy of the communities of a particular size versus the community size (in 
log scale) for PhoNet and $PhoNet_{rand}$.} 
\label{entropy}
\end{figure*}

Figure~\ref{entropy} illustrates, for all the communities obtained from the clustering of PhoNet, the average feature 
entropy exhibited by the communities of a particular size\footnote{Let there be $n$ communities of a particular size 
$k$ picked up at various thresholds. The average feature entropy of the communities of size $k$ is therefore 
$\frac{1}{n}{\sum_{i=1}^n{F_{E_i}}}$ where $F_{E_i}$ signifies the feature entropy of the $i^{th}$ community.} 
(y-axis), versus the community size in log scale (x-axis). 

We next investigate whether or not the communities obtained from PhoNet are better in terms of feature entropy than 
they would have been, if the consonant inventories had evolved just by chance. For this purpose we construct a random 
version of PhoNet and call it PhoNet$_{rand}$. 
\paragraph {Construction of PhoNet$_{rand}$:} For each consonant $c$ let the frequency of occurrence in UPSID$_{317}$ 
be denoted by $f_c$. Let there be 317 bins each corresponding to a language in UPSID$_{317}$. $f_c$ bins are then 
chosen uniformly at random and the consonant $c$ is packed into these bins. Thus the consonant inventories of the 317 
languages corresponding to the bins are generated. PhoNet$_{rand}$ can be then constructed from these new consonant 
inventories similarly as PhoNet. The method is summarized in Algorithm~\ref{rand}.

\protect\begin{algorithm}
\BlankLine
 \For{each consonant c}
  {
  \For{i = 1 to $f_c$}
   {
    Choose one of the 317 bins, corresponding to the languages in UPSID$_{317}$, uniformly at random;
	\BlankLine
	Pack the consonant $c$ into the bin so chosen if it has not been already packed into this bin earlier;  
   } 
  } 
 \BlankLine
 Construct PhoNet$_{rand}$, similarly as PhoNet, from the new consonant inventories (each bin corresponds to a new 
inventory);   
\label{rand}
\caption{Algorithm to construct PhoNet$_{rand}$}    
\end{algorithm}     

We apply Algorithm~\ref{wght} in order to find the communities appearing in PhoNet$_{rand}$. The average feature 
entropy for the communities of a particular size (y-axis), versus the community size in log scale (x-axis) are shown 
in Figure~\ref{entropy} (along with the curve for PhoNet). A closer inspection of the curves immediately makes it 
clear that the average feature entropy exhibited by the communities of PhoNet are substantially better than that of 
PhoNet$_{rand}$ especially when the community size remains less than 15. As this size increases, the difference in the 
average feature entropy of the communities of PhoNet and PhoNet$_{rand}$ gradually diminishes. This is because, the 
community then comprises of almost all the nodes of PhoNet which are also the nodes of PhoNet$_{rand}$. Hence the 
average feature entropy exhibited by the respective giant components of PhoNet and PhoNet$_{rand}$ is close and this 
closeness increases with the increase in the size of the giant component.   
 
\begin{figure*}
\centerline{\psfig{file=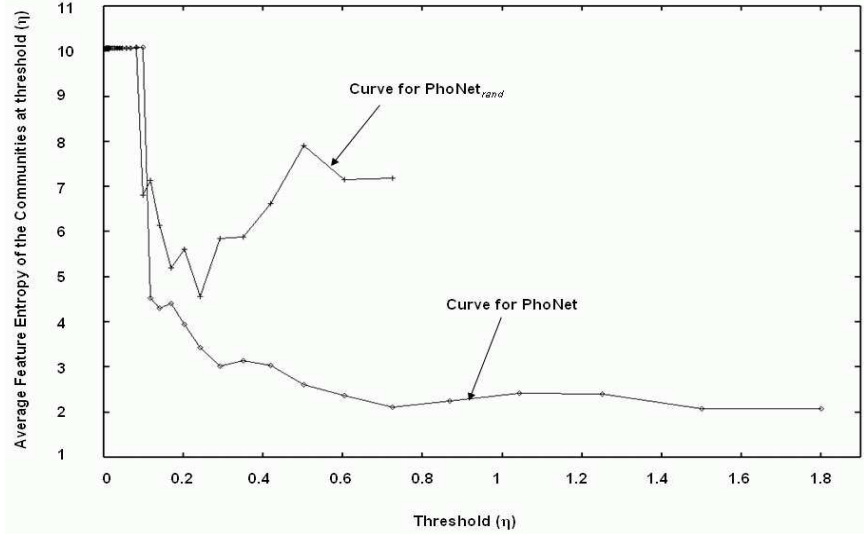,width=4.5in}}
\caption{Average feature entropy of the communities obtained at threshold $\eta$ versus the threshold $\eta$ for both 
PhoNet and PhoNet$_{rand}$} 
\label{thravg}
\end{figure*}

Figure~\ref{thravg} (showing the average feature entropy of the communities for different values of $\eta$ in the 
y-axis versus the threshold $\eta$ in the x-axis) further strengthens the fact that feature entropy exhibited by the 
communities occurring in PhoNet are substantially better than those occurring in PhoNet$_{rand}$. It clearly shows 
that the average feature entropy of the communities, obtained at all thresholds greater than 0.2, are significantly 
lower in case of PhoNet than in PhoNet$_{rand}$. Below this threshold, gradually the average feature entropy of the 
communities of PhoNet and PhoNet$_{rand}$ come closer, until they are identical. Another important observation is that 
the communities of PhoNet$_{rand}$ do not emerge at thresholds greater than 0.8. This points to the fact that strong 
patterns of co-occurrence would not have surfaced if the consonant inventories had just evolved by chance.  

The above results not only validate our definition of feature entropy but is also indicative of the fact that the 
community structures observed in PhoNet are not arbitrary and are true representatives of feature economy claimed to 
be observed across languages. In fact, the argument can be further validated by looking into the languages recorded in 
UPSID$_{317}$ and examining whether or not the consonants forming the communities in PhoNet occur in these languages 
so as to minimize feature entropy. 

\begin{figure*}
\centerline{\psfig{file=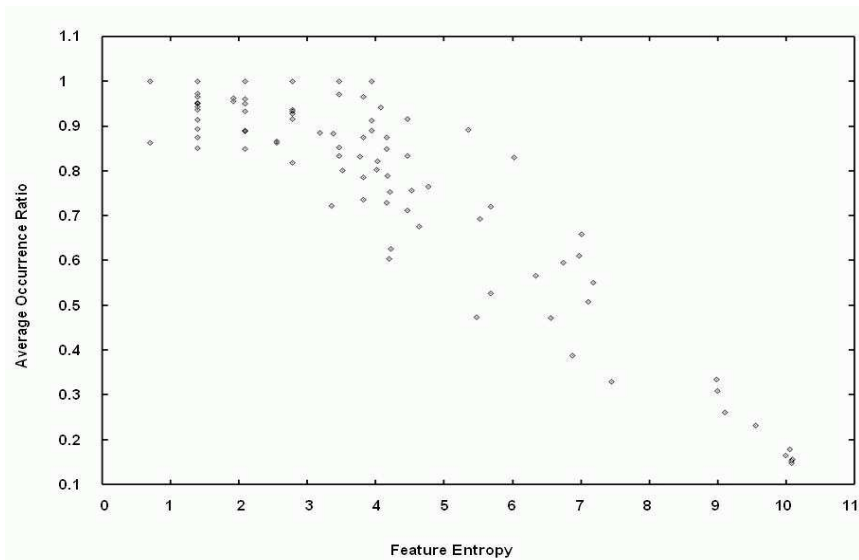,width=4.5in}}
\caption{Average occurrence ratio ($O_{av}$) versus the feature entropy of the communities. Each point corresponds to 
a single community} 
\label{back_st}
\end{figure*}

Figure~\ref{back_st} shows the scatter plot of the average occurrence ratio of the communities obtained from PhoNet 
(y-axis) versus the feature entropy of these communities (x-axis). Each point in this plot corresponds to a single 
community. The plot clearly indicates that the communities exhibiting lower feature entropy have a higher average 
occurrence ratio. For communities having feature entropy less than or equal to 3 the average occurrence ratio is never 
less than 0.7 which means that the consonants forming these communities occur together on an average in 70\% or more 
of the world's languages. As feature entropy increases this ratio gradually decreases until it is almost close to 0 
when feature entropy is around 10. This again attests the fact that the driving force behind the formation of these 
communities is the principle of feature economy and languages indeed tend to choose consonants in order to maximize 
the use of the distinctive features, which are already available in their inventory.    

\section{Conclusions and Discussions}\label{conc}

In this paper we have explored the co-occurrence principles of the consonants, across the inventories of the world's 
languages. Firstly, we have presented an automatic procedure to capture the co-occurrence patterns of the consonants 
across languages. It is important to mention here that this automation also provides an algorithmic definition of {\em 
natural classes}~\cite{Chomsky:68} of phonemes (Table~\ref{tab1} is a natural class of plosives). This is significant 
because there is no single satisfactory definition of such natural classes in literature~\cite{Flemming:05}. The 
communities that we obtained from PhoNet are such natural classes and we can derive them just by regulating the 
threshold of our algorithm.      

Secondly, in order to quantify feature economy we have introduced the notion of feature entropy. This quantification 
immediately allows us to validate the explanation of the organizational principles of the sound inventories in terms 
of feature economy, provided by the earlier researchers.   

Some of our important findings from this work are,
\begin{itemize}
\item The patterns of co-occurrence of the consonants, reflected through communities in PhoNet, are observed in 80\% 
or more of the world's languages;
\item Such patterns of co-occurrence would not have emerged if the consonant inventories had evolved just by chance;
\item The consonant communities that maximize feature economy tend to occur more frequently (70\% or higher number of 
times) in the languages of the world.
\end{itemize}

Until now we have emphasized on the fact that feature economy is the driving force behind the formation of consonant 
communities. An issue which draws attention is that how such a force might have originated. One possible reason could 
be due to certain general principles like {\em maximal perceptual contrast}~\cite{Lindblom:88} and {\em articulatory 
ease}~\cite{Boer:00,Lindblom:88} and {\em ease of learnability}~\cite{Boer:00}. For instance, maximal perceptual 
contrast, which is desirable between the phonemes of a language for proper perception of each individual phoneme in a 
noisy environment, would try to reduce feature economy (since better perception calls for use of a larger number of 
distinctive features). On the other hand, ease of learnability, which is required so that a speaker can learn a 
language with minimum effort, tries to increase feature economy (since learnability increases if there are only a few 
distinctive features to be learnt). It would be interesting to see how the quantification of feature economy can help 
us in understanding the interplay, between these principles, that goes on in shaping the structure of the consonant 
inventories. We look forward to do the same as a part of our future work.          

%
%
%
%
%

\end{document}